\documentclass[journal]{IEEEtran}
 \pdfoutput=1
 \usepackage{cite}
\usepackage{amsmath,amssymb,amsfonts}
\usepackage{algorithmic}
\usepackage{graphicx}
\usepackage{textcomp}
\usepackage{framed,multirow}
\usepackage{latexsym}
\usepackage{amsmath}
\usepackage{threeparttable}
\usepackage{float}
\usepackage{array}
\usepackage{hhline}
\usepackage{colortbl}
\usepackage{stfloats}
\usepackage{booktabs}
\usepackage{multirow}
\usepackage{enumerate}
\usepackage{diagbox}
\usepackage{url}
\usepackage{hyperref}
\usepackage{makecell,rotating}

\def\etal{\textit{et~al}.}

\DeclareGraphicsExtensions{.pdf,.jpeg,.png,.jpg}

\graphicspath{{fig/}}

\definecolor{mygray}{gray}{.9}

\definecolor{newcolor}{rgb}{.8,.349,.1}

\makeatletter

\newcommand{\Rmnum}[1]{\expandafter\@slowromancap\romannumeral #1@}
\makeatother

{Shell \MakeLowercase{\textit{Lin et al.}}: PDBL: Improving Histopathological Tissue Classification with Plug-and-Play Pyramidal Deep-Broad Learning}

\title{PDBL: Improving Histopathological Tissue Classification with Plug-and-Play Pyramidal Deep-Broad Learning}

\author{Jiatai Lin$^\dagger$, Guoqiang Han$^\dagger$, Xipeng Pan$^\dagger$, Hao Chen, Danyi Li, Xiping Jia, Zhenwei Shi, Zhizhen Wang, Yanfen Cui, Haiming Li, Changhong Liang, Li Liang, Zaiyi Liu, Chu Han
\thanks{This work was supported by the Key R\&D Program of Guangdong Province, China (No. 2021B0101420006), the National Science Fund for Distinguished Young Scholars (No.81925023), the National Natural Science Foundation of China (No. 62102103, 81771912, 82071892, 82001789, 81901704, 81702322, 81472712, 81772918, 81972277 and 62002082), High-level Hospital Construction Project (No. DFJH201805 and DFJH201914), Natural Science Foundation of Guangdong Province (No. 2017A030313896), Guanzhou Science \& Technology Project (No. 201804010319). (Corresponding author: Li Liang, Zaiyi Liu, Chu Han.)}
\thanks{Jiatai Lin and Guoqiang Han are with the School of Computer Science and Engineering, South China University of Technology, Guangzhou, China.}
\thanks{Chu Han, Xipeng Pan, Zhenwei Shi, Changhong Liang and Zaiyi Liu are with the Department of Radiology, Guangdong Provincial People’s Hospital, Guangdong Academy of Medical Sciences, Guangzhou, Guangdong, 510080, China. E-mail: hanchu@gdph.org.cn.}
\thanks{Hao Chen is with the Department of Computer Science and Engineering, The Hong Kong University of Science and Technology, Clear Water Bay, Hong Kong.}
\thanks{Xiping Jia is with Guangdong Polytechnic Normal University, School of Computer Science, Guangzhou, China, 510665.}
\thanks{Li Liang and Danyi Li are with Department of Pathology, Nanfang Hospital and Basic Medical College, Southern Medical University, Guangzhou 510515, Guangdong, China.
Guangdong Province Key Laboratory of Molecular Tumor Pathology, Guangzhou 510515, Guangdong, China.}
\thanks{Haiming Li is with the Department of Radiology, Fudan University Shanghai Cancer Center, Shanghai 200032, China and the Department of Oncology, Shanghai Medical College, Fudan University, Shanghai 200032, China.}
\thanks{Yanfen Cui is with the Department of Radiology, Shanxi Province Cancer Hospital, Taiyuan, Shanxi, 030013, China.}
\thanks{$^\dagger$The first three authors contribute equally.}
}

\begin{document}
\maketitle

\IEEEtitleabstractindextext{\begin{abstract}
Histopathological tissue classification is a fundamental task in pathomics cancer research. Precisely differentiating different tissue types is a benefit for the downstream researches, like cancer diagnosis, prognosis and etc. Existing works mostly leverage the popular classification backbones in computer vision to achieve histopathological tissue classification. In this paper, we proposed a super lightweight plug-and-play module, named Pyramidal Deep-Broad Learning (PDBL), for any well-trained classification backbone to further improve the classification performance without a re-training burden. We mimic how pathologists observe pathology slides in different magnifications and construct an image pyramid for the input image in order to obtain the pyramidal contextual information. For each level in the pyramid, we extract the multi-scale deep-broad features by our proposed Deep-Broad block (DB-block). We equipped PDBL in three popular classification backbones, ShuffLeNetV2, EfficientNetb0, and ResNet50 to evaluate the effectiveness and efficiency of our proposed module on two datasets (Kather Multiclass Dataset and the LC25000 Dataset). Experimental results demonstrate the proposed PDBL can steadily improve the tissue-level classification performance for any CNN backbones, especially for the lightweight models when given a small among of training samples (less than 10\%), which greatly saves the computational time and annotation efforts.
\end{abstract}
}

\maketitle
\IEEEdisplaynontitleabstractindextext

\IEEEpeerreviewmaketitle

%!Tex root=./tmi.tex
\section{Introduction}
Histopathological slides not only play a vital role in cancer diagnosis, but also deliver valuable tumor microenvironment information for cancer research~\cite{campanella2019clinical,kather2019deep}. To analyze the whole slide images by computer algorithms is crucial for precision medicine on cancers, such as diagnosis prediction~\cite{coudray2018classification, gehrung2021triage}, molecular status prediction~\cite{binder2021morphological,fu2020pan} and even the origins of the unknown primary cancers prediction~\cite{lu2021ai}. Segmenting and recognizing various tissue types is the very first step of histopathological image analysis. Semantic segmentation~\cite{van2021hooknet} is the best way to define tissue types for every single pixel. However, due to the gigapixel resolution and the expertise requirement, obtaining pixel-level annotations is extremely difficult and time-consuming~\cite{srinidhi2021deep}. Therefore, patch-level classification now becomes an alternative solution~\cite{xue2021selective,kather}, which can greatly save the annotation efforts.

Convolutional neural network (CNN) has demonstrated outstanding performance in image classification problem~\cite{deng2009imagenet}, with a series of classification backbones, i.e., ResNet~\cite{he2016deep}, ShuffLeNet~\cite{zhang2018shufflenet}, Efficient-Net~\cite{backbone_eff} and etc. They have been rapidly extended to medical image classification~\cite{zhang2019medical,cai2020review}, including histopathological image classification~\cite{gecer2018detection,wang2019rmdl}. Typically, Han~\etal~\cite{2017Breast} presented a CNN-based multi-classification model for histopathological tissue classification of breast cancer, which achieves over 94\% patch-level accuracy under four magnification factors. Tsai~\etal~\cite{tsai2021deep} tested five common classification backbones on colorectal tissue classification using 100,000 training image patches. All the backbones can achieve over 95\% accuracy. The above researches prove that the current CNN classification backbones have already demonstrated strong feature representation ability and achieved promising results for histopathology tissue classification. In this paper, we reconsider how to make good use of the features extracted from the existing CNN backbones, to further improve the classification performance as well as to increase the model generalizability, adaptability, and robustness.

In clinical practice, pathologists read histopathological slides by switching the object lens to observe the slides under different magnifications. Therefore, considering multi-scale contextual information is critical for histopathological image analysis. Inspired by this observation, we proposed a lightweight plug-and-play module for any CNN classification backbones, named Pyramidal Deep-Broad Learning (PDBL). An image pyramid is constructed to extract the pyramidal contextual information. For each level in the pyramid, we proposed a Deep-Broad block (DB-block) to fully discover the multi-scale deep-broad features extracted by the CNN backbones from low level to high level. Our proposed PDBL can be plugged on any classification backbone and effectively improves the classification performance with very few extra computational resources.

We tested PDBL on three representative CNN backbones, ShuffLeNetV2, EfficientNetb0, and ResNet50 on Kather Multiclass Dataset~\cite{kather} and Lung Colon Cancer Histopathological Image Dataset~\cite{LC25000}. We conduct two main experiments, one is the effectiveness of PDBL with different proportions of the training samples, the other is the robustness of PDBL with only $1\%$ training samples. Experimental results demonstrate that PDBL effectively improves the classification performance on both datasets. When very limited training samples are involved (1\% only), PDBL can maintain a standout improvement compared with the baseline models without PDBL. It can significantly reduce the annotation efforts and computational resources. Experimental results also show that PDBL improves domain adaptation abilities for CNN models. The contributions of this paper are summarized as follows:

\begin{itemize}
  \item[$\bullet$] We proposed a lightweight plug-and-play module (PDBL), which can be easily applied on almost any common CNN-based classification backbone. It can generally improve all the three CNN backbones we have tested for histopathological tissue classification with no re-training burden.
  \item[$\bullet$] We proposed a Deep-Broad block to fully discover the multi-scale deep-broad features from low level to high level.
  \item[$\bullet$] The proposed PDBL demonstrates outstanding improvement of the performance for the lightweight models with very limited training samples (1\% only).
  \item[$\bullet$] Models with PDBL can relieve the requirement of large scale training data and be easily and efficiently adapted to a new domain with only a few training samples, which greatly saves the computational resources and annotation efforts.
\end{itemize} 
%!Tex root=./tmi.tex
\section{Related Works}\label{sec2}
\subsection{Histopathological Image Classification}
%Histopathological image classification can be roughly divided into tissue-level classification and slide-level classification.
Automated analysis of whole slide images (WSIs) plays a crucial role in computer-assisted tumor diagnosis~\cite{coudray2018classification,hollon2020near}. Due to the giga-pixel resolution, directly processing the entire WSI is not feasible. Moreover, obtaining pixel-level annotation is extremely difficult. Hence, histopathological tissue classification has been widely employed as an alternative solution for tissue semantic segmentation of WSIs~\cite{kather}.

With the development of CNN models, most of the histopathological image classification models~\cite{hou2016patch,xu2017large} are originated from the popular classification backbones from the natural image classification. However, histopathological image classification faces different challenges, such as extremely large image resolution, deficiency of labels and multi-scale information integration~\cite{2021Role}. WSI-Net~\cite{ni2019wsi} model was proposed to add an additional classification branch to discard the normal tissue in order to save computational resources. Raczkowski~\etal~\cite{rkaczkowski2019ara} proposed a pathologist-in-the-loop model to solve the insufficient labeling problem. Xue~\etal~\cite{xue2021selective} proposed to synthesis histopathological patch images using GAN to enhance the feature representation and improve the classification performance. Many studies attempted to extract multi-scale features to better solve the classification problem of histopathological images with end-to-end deep learning models, such as the Deep-Hipo model~\cite{kosaraju2020deep} and multi-resolution model~\cite{li2021multi}.

In this paper, we used a broad learning strategy to fully discover the deep features extracted by deep learning models and leverage the multi-scale contextual information to improve the performance of the CNN-based models without excessive computational costs. Next, we will introduce some common deep learning architectures and broad learning system approaches.

\subsection{Deep Learning Architectures}
Deep learning models have already dominated the image classification problem~\cite{goodfellow2016deep}. They usually go through several stages to reduce the feature dimension and to extract higher-level semantic features, defined in Eq.~\ref{cnn}.
\begin{equation}\label{cnn}
\mathcal{F} = \mathbf{stage}_{1}\odot \mathbf{stage}_{2}\odot ... \odot \mathbf{stage}_{j}(X),j=1,2,...,h.
\end{equation}
where $\mathcal{F}$ denotes the CNN model and $X$ is the input image. Each stage is composed of a series of cascaded convolutional blocks, such as Res-block~\cite{he2016deep}, Efficient-block~\cite{backbone_eff}, Shuffle-block~\cite{ma2018shufflenet} and Inception-block~\cite{szegedy2017inception}. These blocks were designed to prevent the gradient vanishing problem and to increase the capacity of CNN models by balancing the depth and width of the deep architecture. Some skip connections were also introduced to transmit the features between different convolutional layers to avoid information loss and enhance the feature representation.

The current CNN classification backbones have already had strong capacity and feature representation ability. So in this paper, we aim to discover the potential of the multi-scale deep features extracted from different stages, and further improve the performance for any well-trained CNN backbones for histopathological tissue classification.
\begin{figure*}
  \centering
  \includegraphics[width=.99\linewidth]{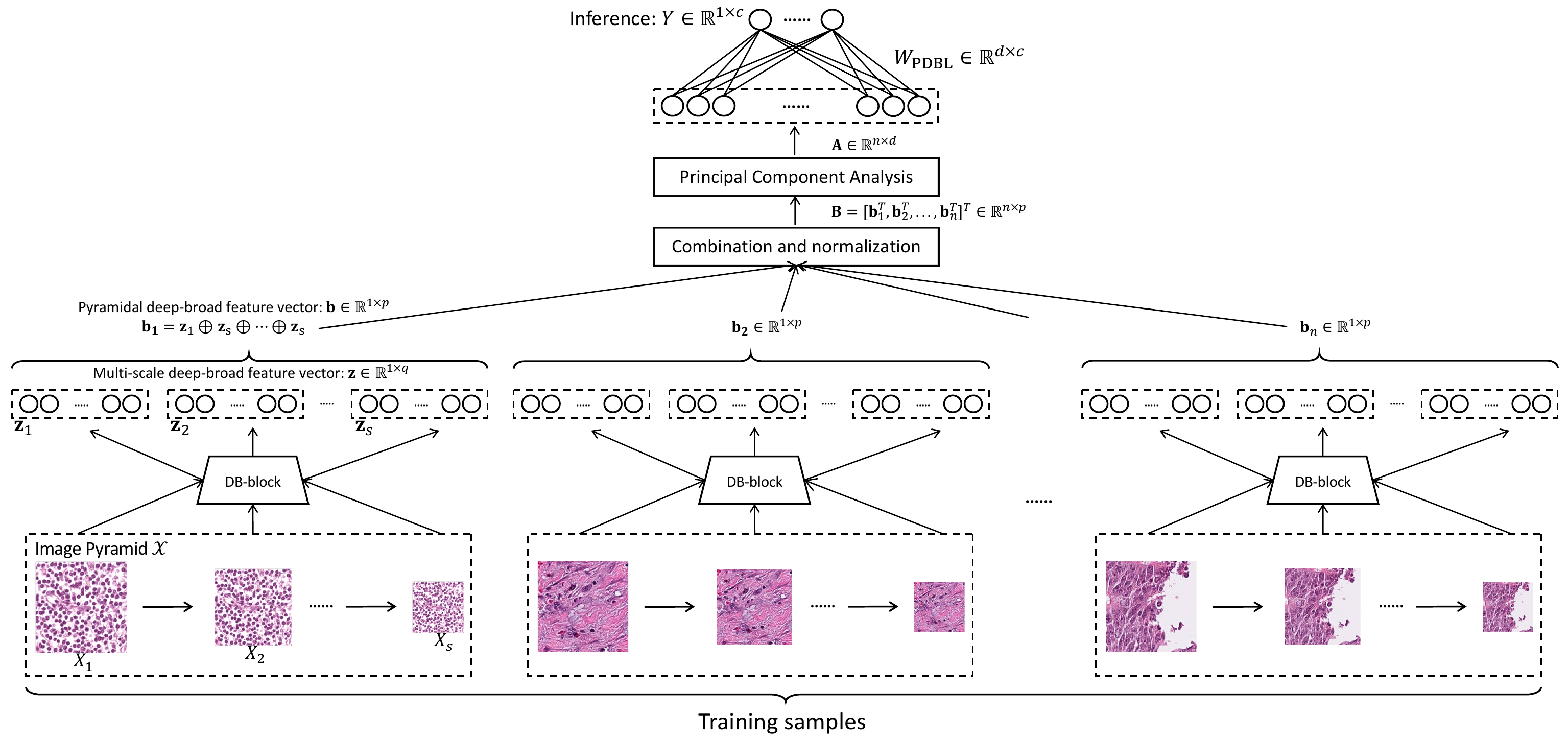}
  \caption{Overview of the proposed Pyramidal Deep-Broad Learning (PDBL). In PDBL, we create an image pyramid for each image in the training set to obtain the pyramidal contextual information. For each image in the pyramid, we extract the multi-scale deep-broad features by a Deep-Broad block (DB-block). Finally, histopathological image categories can be inferred by the broad learning system. For simplification, we only show the notations of the left input image.}
  \label{fig:PDBM}
\end{figure*}
\subsection{Broad Learning System}
With the breakthrough of the GPU architectures over the past decade, researches have kept increasing the depth of CNN models and achieved outstanding performance in most of the computer vision and medical imaging tasks. However, the deeper the network is, the more computational time of the training we spend. Chen~\etal~\cite{BLS} proposed an opposite direction of neural network by expanding the width instead of increasing the depth, called Broad Learning system (BL). BL tends to breadth-wise expand the feature space by multi-group feature mapping, and uses a shallow fully-connection layer to calculate output, which greatly saves the computational resources comparing with deep learning (DL).

In the past years, a series of BL approaches~\cite{UBLS,MBLS,RBLS,FBLS} have been proposed. The motivation of these approaches is to provide an architecture to breadth-wisely combine multiple groups of features by solving the following optimization problems:
\begin{equation}\label{optimization}
  W_{opt} = \mathop{\arg\min}_{W_{init}} \parallel AW_{init} - Y \parallel _2^2 + \gamma \parallel W_{init} \parallel_2^2
\end{equation}
where $Y$, $A$ represent target matrix(vector) and combined feature matrix that concatenates with all groups of feature nodes and enhance nodes. $W_{init}$ and $W_{opt}$ are pre-update and post-update weights of output layer, which can be updated rapidly by pseudo-inverse method:
\begin{equation}\label{Ridge_Regression}
  W_{opt} = A^+ Y = \lim_{\lambda \rightarrow 0}(AA^T + \lambda E)^{-1}A^TY
\end{equation}
where $E$ and $\lambda$ represent an identity matrix and a constant parameter. When $\lambda = 0$, the updating method is ridge regression which requires $A$ to be a non-singular matrix.

In short, DL has stronger semantic feature extraction ability, while BL is faster and more lightweight. So in this paper, we want to gather the strengths of both DL and BL by using DL to extract multi-scale semantic features and using BL for inference. Such deep-broad design is effective and will not introduce extra training burden. Considering histopathological tissue classification, we designed a pyramidal structure for the deep-broad learning to consider the pyramidal contextual information of the histopathological images. 
%!Tex root=./tmi.tex
\begin{table}[t]
  \centering
  \caption{Symbol annotations}\label{symbol}
  \begin{tabular}{c|c}
    \hline
    % after \\: \hline or \cline{col1-col2} \cline{col3-col4} ...
    $\mathbf{Symbol}$    &$\mathbf{Meaning}$                    \\ \hline
    $I$       &Input image (WSI patch)      \\ \hline
    $X$       &Sub-image of the image pyramid      \\ \hline
    $\mathcal{X}$       &Image pyramid       \\ \hline
    $f$     &Intermediate deep features                        \\ \hline
    $\textbf{e}$ &Channel-wise feature vector of $f$      \\ \hline
    $\textbf{z}$   & Multi-scale deep-broad feature vector of $X$     \\ \hline
    $\textbf{b}$   & Pyramidal deep-broad feature vector of  $\mathcal{X}$         \\ \hline
    $\textbf{B}$   & Feature matrix of the complete training data    \\ \hline
    $\textbf{C}$     &Covariance matrix                                \\ \hline
    $\textbf{U}$       &Dimension reduction matrix                       \\ \hline
    $\textbf{A}$       &Feature matrix after PCA         \\ \hline
    $W_{\text{PBDL}}$ &Weights of PDBL             \\ \hline
    $n$     &Number of the training samples \\ \hline
    $c$     &Number of the categories                               \\ \hline
    $q$       &Dimension of $\textbf{z}$  \\ \hline
    $p$       &Dimension of $\textbf{b}$        \\ \hline
    $d$       &Target dimension in dimensionality reduction     \\ \hline

    \hline
  \end{tabular}
\end{table}

\section{Pyramidal Deep-Broad Learning}\label{sec3}

\begin{figure}
  \includegraphics[width=.99\linewidth]{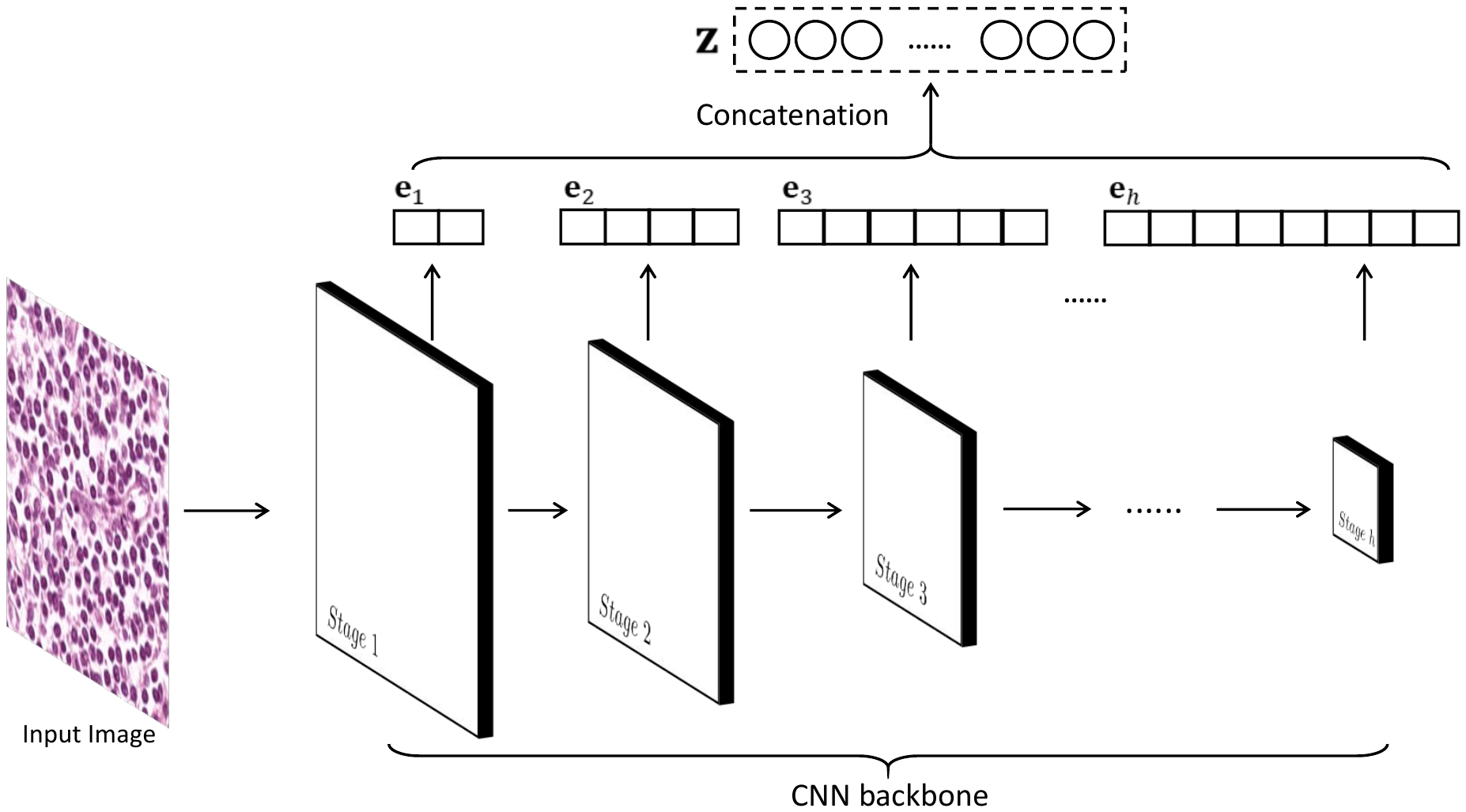}
  \caption{Illustration of DB-block. Given any CNN backbone, intermediate feature maps of each stage are compressed into channel-wise features $\textbf{e}$ by the adaptive global average pooling. Then we can obtain the multi-scale deep-broad features $\textbf{z}$ by concatenating $\textbf{e}$.}
  \label{fig:db-block}
\end{figure}

Deep learning~(DL) has the powerful feature extraction ability while Broad Learning~(BL) is good at combining multiple groups of features for fast inference. Theoretically, associating DL with BL can effectively improve the performance of existing CNN-based models. In this section, we proposed a novel Pyramidal Deep-Broad Learning~(PDBL) with a Deep-Broad block (DB-block) for histopathological tissue classification. Fig.~\ref{fig:PDBM} demonstrates the overview of the proposed PDBL. We first construct an image pyramid for the input image. And then we extract the multi-scale deep-broad features by DB-block, shown in Fig.~\ref{fig:db-block}. Finally, a broad learning system was introduced for the inference. The annotations in this article are defined in Table~\ref{symbol}.

\subsection{Image Pyramid Construction}
Typically, pathologists observe pathological sections under different magnifications. Inspired by this, we want the CNN-based models to be able to consider the multi-scale contextual information too. Given an input WSI patch $I$, we construct an image pyramid by downscaling the input image for $s$ times, defined as follows:
\begin{equation}
X_i = \xi(I,S_i), i=1,...,s\
\label{pyramid_gen}
\end{equation}
where $\xi$ denotes the scale transformation of bilinear interpolation with the scaling factor $S_i$.

Now we have an image pyramid $\mathcal{X}$ with $s$ sub-images, including the input image $X_1=I$.
\begin{equation}
\mathcal{X} = \{X_1, X_2, ... , X_s\}
\label{pyramid}
\end{equation}
And then, each sub-image is passed into our proposed DB-block for feature extraction.

\subsection{Multi-Scale Deep-Broad Feature Extraction via Deep-Broad Block}
\label{sec:DB-block}
We proposed a Deep-Broad block (DB-block) to extract features for the image pyramid $\mathcal{X}$, as shown in Fig.~\ref{fig:db-block}. The DB-block broadens the deep features from each stage of the CNN backbone and forms the multi-scale deep-broad features.

For the sub-image $X_i$, we first extract its intermediate deep features $f$ from the last layer of each stage of the CNN backbones. The intermediate deep features $f_k$ at the $k$-th stage are squeezed into a channel-wise feature vector by adaptive global average pooling in Eq.~\ref{squeeze}.
\begin{equation}\label{squeeze}
\textbf{e}_k = \frac{1}{H_k\times W_k}\sum_{i=1}^{H_k}\sum_{j=1}^{W_k}f_k(i,j)
\end{equation}
where $H_k$ and $W_k$ represent the height and width of the intermediate feature maps at the $k$-th stage.

The multi-scale deep-broad features $\textbf{z}_i$ of the sub-image $X_i$ can be obtained by concatenating all the channel-wise feature vector $\textbf{e}$, as follows:
\begin{equation}\label{concat}
\textbf{z}_i = \textbf{e}_{1} \oplus \textbf{e}_{2} \oplus ,..., \oplus \textbf{e}_{h}
\end{equation}
where $\textbf{z}_i \in \mathbb{R}^{1\times q}$, and $q$ represents the dimensions of this feature group, which is the summation of channel numbers of all the stages. $h$ denotes the number of stages of the CNN backbone. $\oplus$ represents the concatenation operation.

Now for the image pyramid $\mathcal{X} = \{X_1, X_2, ... , X_s\}$, we can obtain pyramidal deep-broad feature vector $\textbf{b}$ with pyramidal contextual information by concatenating all the multi-scale deep-broad features $\textbf{z}$ of the sub-images as follows:

\begin{equation}\label{broaden_feature_vector}
  \textbf{b} = \textbf{z}_1 \oplus \textbf{z}_2 \oplus,...,\oplus \textbf{z}_s
\end{equation}
where $\textbf{b} \in \mathbb{R}^{1 \times p}$ is the pyramidal deep-broad feature vector of the image pyramid $\mathcal{X}$. And $p$ is the dimension of $\textbf{b}$ where $p = s \times q$.

In DB-block, we extract the deep learning features by the baseline model pre-trained with only train the CNN backbone once using the original training set. And we extract features of all the sub-images from the image pyramid by this CNN backbone.

\subsection{Broad Learning Inference}
With the pyramidal deep-broad feature vector $\textbf{b}$ of the image pyramid $\mathcal{X}$, we apply a broad learning system for inference. Let us denote the complete training samples as $\mathbb{X}=\{{\mathcal{X}_i|i=1,2,...,n}\}$, where $n$ is the number of training samples. We can obtain a set of feature vectors $\{\textbf{b}_i|i=1,2,...,n\}$. Then a broad feature matrix $\textbf{B}$ is constructed by combining all the feature vectors $\textbf{b}_i$ in Eq.~\ref{norm}.
\begin{equation}\label{norm}
  \textbf{B} = \delta([\textbf{b}_1^\mathsf{T},\textbf{b}_2^\mathsf{T},...,\textbf{b}_n^\mathsf{T}]^\mathsf{T}), \textbf{B}\in \mathbb{R}^{n\times p}
\end{equation}
where $\delta$ denotes the matrix normalization transformation.

%At first, multiple broad feature vectors are concatenated and reduced in dimension to feature matrix $A$. Furthermore, a weight-based full-connected layer is used to connect features for final inference. To rapidly updating, the pseudo-inverse algorithm is employed to obtain the optimal weight $W_{bls}$. The details are shown in the following parts.

%\textbf{Principal component analysis:}
In order to reduce the feature dimension and redundancy, principal component analysis~(PCA) is employed for $\textbf{B}$. We first calculate the covariance matrix $\textbf{C}$ by Eq.~\ref{cov}:
\begin{equation}\label{cov}
  \textbf{C} = \frac{1}{n}\textbf{B}^\mathsf{T}\textbf{B}, \textbf{C}\in \mathbb{R}^{p\times p}.
\end{equation}
Furthermore, dimension reduction matrix is obtained by SVD algorithm:
\begin{equation}\label{svd}
  [\textbf{U},\Sigma, \textbf{V}^\mathsf{T}] = SVD(\textbf{C}).
\end{equation}
where $\textbf{U}\in\mathbb{R}^{p\times d}$, and $d$ represents the target dimension in dimensionality reduction.

According to $\textbf{U}$, the dimension of feature matrix $\textbf{B}$ can be reduced to the matrix $\textbf{A}$ by Eq.~\ref{dr}:
\begin{equation}\label{dr}
  \textbf{A} = \textbf{B}\times \textbf{U}, \textbf{A}\in \mathbb{R}^{n\times d}.
\end{equation}

%\textbf{Pseudo-inverse algorithm:}
Finally, the probabilities $Y$ of all the categories can be calculated by:
\begin{equation}\label{output}
  Y = \textbf{A} W_{\text{PDBL}}, W_{\text{PDBL}} \in \mathbb{R}^{d \times c}
\end{equation}
where $W_{\text{PDBL}}$ is the weights of PDBL, which can be calculated by the pseudo-inverse algorithm. $c$ denotes the number of categories.

\textbf{In the training phase,} we use $\textbf{A}_{train}$ to represent the feature matrix for the complete training set after PCA. According to the ground truth labels $Y_{train}$, the weights $W_{\text{PDBL}}$ of PDBL can be calculated by the pseudo-inverse algorithm as follows:
\begin{equation}\label{output}
  W_{\text{PDBL}} = \textbf{A}_{train}^+ Y_{train}
\end{equation}
where $\textbf{A}_{train}^+$ can be calculated by:
\begin{equation}\label{calculation_A}
  \textbf{A}_{train}^+ = \mathbf{lim}_{\lambda\rightarrow 0}(\lambda E+\textbf{A}_{train}\textbf{A}_{train}^T)^{-1}\textbf{A}_{train}^T
\end{equation}
where $\lambda$ and $E$ represent a constant and unit matrix, respectively. Since pseudo-inverse algorithm only updates the weights for once, it greatly saves the computational resources.

\textbf{In the testing phase}, we can obtain the feature matrix $\textbf{A}_{test}$ by the same steps of the training phase. Then we can infer the probabilities $Y_{test}$ of tissue categories by:
\begin{equation}\label{Infer}
  Y_{test} = \textbf{A}_{test}W_{\text{PDBL}}
\end{equation}

The final classification results is the tissue categories with the largest probabilistic value.

Since the proposed PDBL is a plug-and-play module. It can be applied to any CNN backbone and further improve the classification performance.

\subsection{Implementation and Training Details}
In our experiments, all the CNN backbones were implemented in PyTorch. The backbones were trained on an NVIDIA RTX 2080Ti with the cross entropy loss and the SGD optimizer with $1e-3$ learning rate and $0.9$ momentum. The batchsize was set to $20$. Patches were resized into $224\times 224$ and normalized in both the training and test phase. All the backbones were fine-tuned for $50$ epochs with weights provided by PyTorch which were pretrained on ImageNet.

For our proposed PDBL, we created image pyramid under three different resolution ($112\times 112$, $160 \times 160$ and $224 \times 224$). The deep-broad features of the images of different scales were all extracted from the network trained by the original image size ($224 \times 224$). And the target dimension $d$ of PCA was decided by the total number $n$ of the training samples as follows:
\begin{equation}
    b = \left\{
               \begin{array}{cc}
                 0.9*n, & n<=2000 \\
                 2000, & n>2000
               \end{array}
    \right.
\end{equation}
where $n$ is the number of training samples.

%!Tex root=./tmi.tex

\section{Datasets}\label{sec4}
We evaluate our proposed PDBL in the following two datasets Kather Multiclass Dataset~\cite{kather} and LC25000 Dataset~\cite{LC25000}.
\subsection{Kather Multiclass Dataset}\label{kather_dataset}
Kather Multiclass Dataset is a multi-class colorectal dataset composed of H\&E stained histopathological tissue patches, which was published by J. N. Kather\footnotemark[2]. Kather Multiclass Dataset is composed of two subsets. Kather~\etal~\cite{kather} manually delineated tissue regions in $86$ colorectal~(CRC) tissue slides and they extracted $100,000$ H\&E histopathological tissue patches from these regions as Kather Multiclass Internal~(KMI) subset at $20\times$ magnification. They also extracted an additional independent Kather Multiclass External~(KME) subset include $7180$ H\&E stained histopathological patches. Histopathological images in Kather Multiclass Dataset are cropped to a square size of $224\times 224$ pixels from primeval whole slide images~(WSIs) at $20\times$ magnification.

\begin{figure}[t]
    \setlength{\tabcolsep}{1pt}
    \centering
    \begin{tabular}{ccccc}
        \includegraphics[width=.195\linewidth]{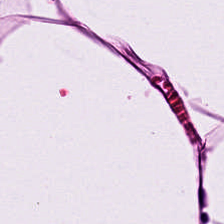}&
        \includegraphics[width=.195\linewidth]{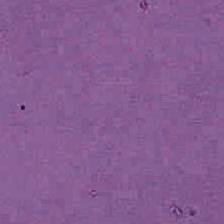}&
        \includegraphics[width=.195\linewidth]{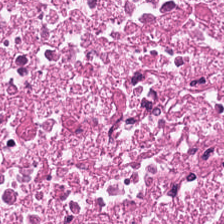}&
        \includegraphics[width=.195\linewidth]{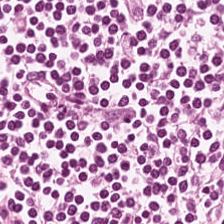}&
        \includegraphics[width=.195\linewidth]{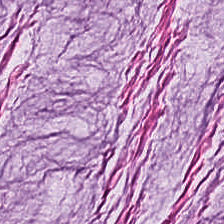}\\
        ADI & BACK & DEB & LYM & MUC \\
    \end{tabular}
    \begin{tabular}{cccc}
        \includegraphics[width=.195\linewidth]{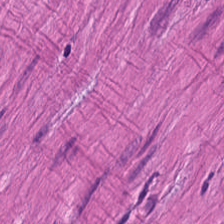}&
        \includegraphics[width=.195\linewidth]{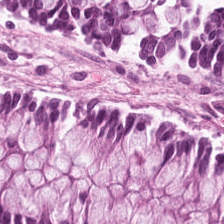}&
        \includegraphics[width=.195\linewidth]{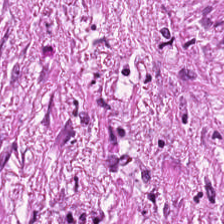}&
        \includegraphics[width=.195\linewidth]{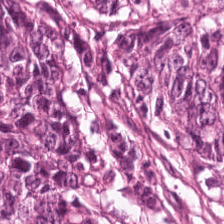}\\
        MUS & NORM & STR & TUM\\
    \end{tabular}
    \caption{Kather Multiclass Dataset includes Adipose~(ADI), background~(BACK), debris~(DEB), lymphocytes~(LYM), mucus~(MUC), smooth muscle~(MUS), normal colon mucosa~(NORM), cancer-associated stroma~(STR), colorectal adenocarcinoma epithelium~(TUM).}
    \label{kather_fig}
\end{figure}

As shown in Fig.~\ref{kather_fig}, each histopathological image belongs to one category of tissues and there are $9$ categories of tissues in Kather Multiclass Dataset such as adipose (ADI), background (BACK), debris (DEB), lymphocytes (LYM), mucus (MUC), smooth muscle (MUS), normal colon mucosa (NORM), cancer-associated stroma (STR), and colorectal adenocarcinoma epithelium (TUM).

\footnotetext[2]{\url{https://zenodo.org/record/1214456}}

\subsection{LC25000 Dataset}\label{lcdataset}
\begin{figure}[t]\centering
    \setlength{\tabcolsep}{1pt}
    \begin{tabular}{ccccc}
        \includegraphics[width=.195\linewidth]{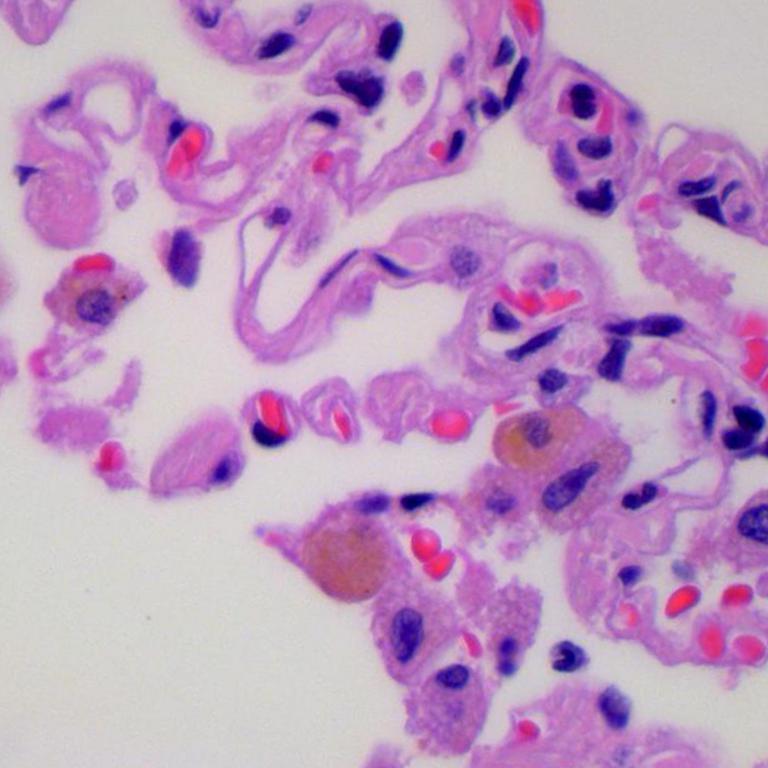}&
        \includegraphics[width=.195\linewidth]{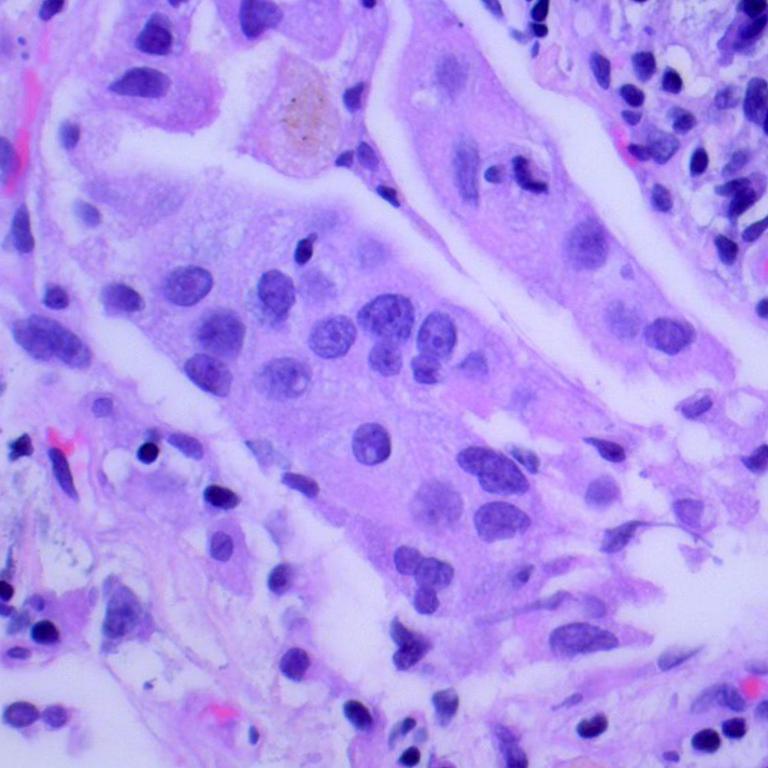}&
        \includegraphics[width=.195\linewidth]{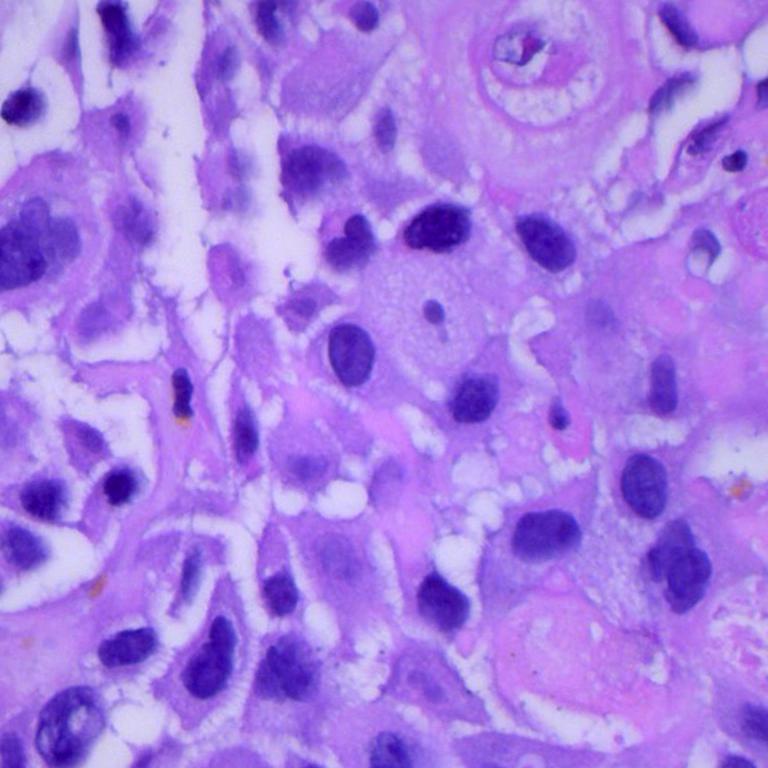}&
        \includegraphics[width=.195\linewidth]{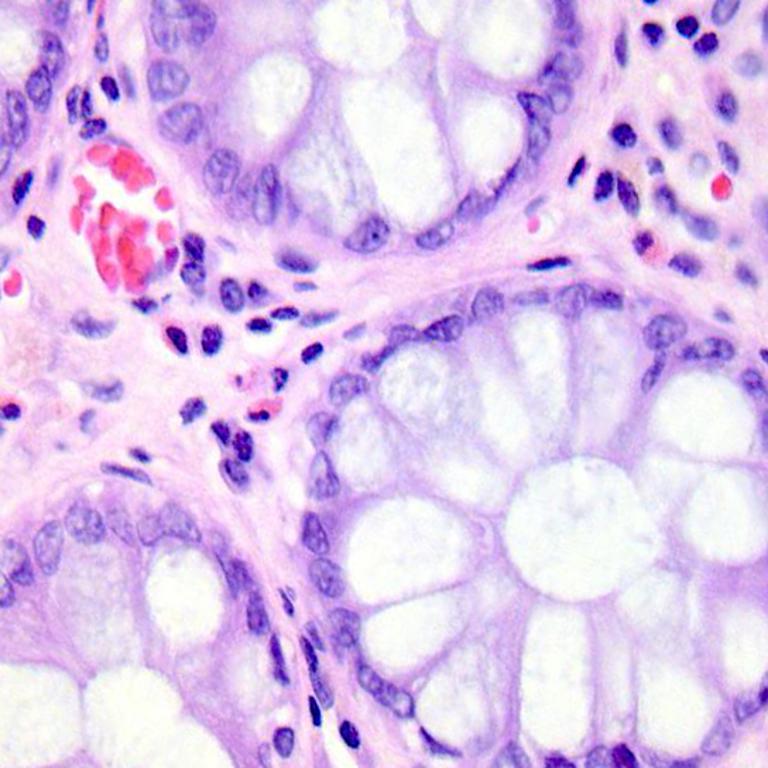}&
        \includegraphics[width=.195\linewidth]{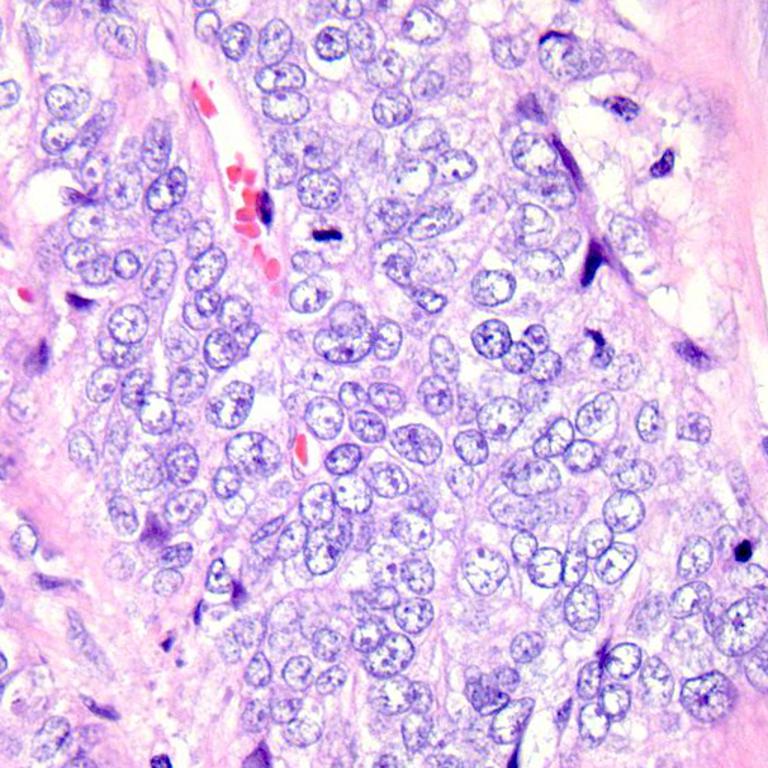}\\
        LN & LAC & LSCC & CN & CAC\\
    \end{tabular}
    \caption{LC25000 Dataset has 5 classes which includes benign lung tissues~(LN), lung adenocarcinomas~(LAC), lung squamous cell carcinomas~(LSCC), benign colonic tissues~(CN) and colon adenocarcinomas~(CAC)}
    \label{fig:lc25000dataset}
\end{figure}

To advance computer-aided automated analysis of lung and colon carcinomas, \cite{LC25000} released a lung and colon histopathological image dataset (LC25000 Dataset)\footnotemark[3]. In LC25000 Dataset, histopathological images are cropped to a square size of $768\times 768$ pixels from H\&E stain WSIs of lung carcinoma and colon carcinoma. As shown in Fig.~\ref{fig:lc25000dataset}, LC25000 Dataset has $5$ categories such as benign lung tissues~(LN), lung adenocarcinomas~(LAC), lung squamous cell carcinomas~(LSCC), benign colonic tissues~(CN), and colon adenocarcinomas~(CAC). LC25000 Dataset is a balanced dataset that each class that has $5000$ histopathological images.

\footnotetext[3]{\url{https://github.com/tampapath/lung_colon_image_set}}

\section{Experimental Results}\label{sec5}
In this section, we conducted several studies to evaluate the proposed PDBL. In the following experiments, PDBL was respectively plugged on three common classification architectures, including EfficientNet-b0~\cite{backbone_eff}, ResNet50~\cite{he2016deep} and a lightweight model ShuffleNetV2~\cite{ma2018shufflenet}. In Section~\ref{exp:proportions}, we evaluate the effectiveness of PDBL with different proportions of training samples. In Section~\ref{exp:robustness}, we test the limit of the proposed PDBL by an extremely difficult task by leveraging only $1\%$ training samples to inference the rest of them (99\%). In Section~\ref{exp:ablation}, we conduct an ablation study to verify the effectiveness and the necessity of the pyramidal design. Next, we demonstrate the advantages of rapid domain adaptation on PDBL in Section~\ref{exp:domain}. Finally, we also show the WSI-level semantic segmentation results by stitching the patch-level classification results.

Accuracy and macro F1 score were used to evaluate the patch-level classification performance of the proposed PDBL in all the experiments.
Due to the page limit, in Table~\ref{tab:quantitative-acc} and Table~\ref{tab:robustness-acc}, we only demonstrate the accuracy. F1 scores can be found in the supplementary materials.

\begin{table*}[t]
	\centering
	\caption{Quantitative comparison (ACC) with different proportions of training samples in Kather Dataset~(100\% training samples: 100,000 patches) and LC25000 Dataset~(100\% training samples: 15,000 patches).}
	\begin{tabular}{ll|c|cccccccc}
		\hline
		\multicolumn{2}{c|}{\multirow{2}{*}{Models}}&\multirow{2}{*}{Dataset}&\multicolumn{8}{c}{Accuracy}\\ \cline{4-11}
&&& 1\%& 5\%& 10\%& 25\%& 35\%& 50\%& 70\%& 100\%\\
		\hline
		\multirow{6}{*}{ShuffLeNetV2}&\textit{Baseline+PDBL} &\multirow{3}{*}{Kather}
&0.86880 &0.92242 &0.94095 &0.94666 &0.94568 &0.94791 &0.95139 &0.95097\\
		&\textit{Baseline*} &
&0.17187 &0.73607 &0.91448 &0.94805 &0.96058 &0.96114 &0.95097 &0.95195\\
		&\textit{Baseline*+PDBL} &
&\textbf{0.87869} &\textbf{0.93816} &\textbf{0.95696} &\textbf{0.96114} &\textbf{0.96365} &\textbf{0.96253} &\textbf{0.96198} &\textbf{0.96156}\\
		\cline{2-11}

		&\textit{Baseline+PDBL} &\multirow{3}{*}{LC25000}
&0.93260  &\textbf{0.96730}   &\textbf{0.97070}    &0.98670 &0.98950    &0.99300    &0.99320    &0.99400\\
		&\textit{Baseline*} &
&0.39280    &0.62380    &0.78980    &0.96810 &0.97310    &0.98480    &0.98890    &0.99660\\
		&\textit{Baseline*+PDBL} &
&\textbf{0.94140} &0.95970 &0.96540 &\textbf{0.98990} &\textbf{0.99140} &\textbf{0.99370} &\textbf{0.99520} &\textbf{0.99720}\\
		\hline
		
		\multirow{6}{*}{EfficientNetb0}&\textit{Baseline+PDBL}
&\multirow{3}{*}{Kather}&0.86685 &0.90933 &0.92214 &0.92841 &0.93064 &0.93036 &0.93287 &0.93398\\
		&\textit{Baseline*} &
&0.91267 &0.93788 &0.94109 &0.94011 &0.94415 &0.93816 &0.93524 &0.94471\\
		&\textit{Baseline*+PDBL} &
&\textbf{0.92256} &\textbf{0.94930} &\textbf{0.94847} &\textbf{0.94930} &\textbf{0.95432} &\textbf{0.95557} &\textbf{0.95641} &\textbf{0.96086}\\
		\cline{2-11}

		&\textit{Baseline+PDBL} &\multirow{3}{*}{LC25000}
&0.94170    &0.97290    &0.98200    &0.99000 &0.99360    &0.99450    &0.99530    &0.99500\\
		&\textit{Baseline*} &
&0.87220    &0.96320    &0.98100    &0.98960 &0.99230    &0.99630    &0.99780    &0.99870\\
		&\textit{Baseline*+PDBL} &
&\textbf{0.95300} &\textbf{0.97990} &\textbf{0.98750} &\textbf{0.99460} &\textbf{0.99610} &\textbf{0.99710} &\textbf{0.99860} &\textbf{0.99940}\\
		\hline

		\multirow{6}{*}{ResNet50}& \textit{Baseline+PDBL} &\multirow{3}{*}{Kather}
&0.86797 &0.92159 &0.93315 &0.93538 &0.93788 &0.93538 &0.93858 &0.93663\\
		&\textit{Baseline*} &
&0.93844 &0.95682 &0.94930 &0.94582 &0.95153 &0.95850 &0.95292 &0.95432\\
		&\textit{Baseline*+PDBL} &
&\textbf{0.93928} &\textbf{0.95933} &\textbf{0.95752} &\textbf{0.95543} &\textbf{0.96212} &\textbf{0.96253} &\textbf{0.96142} &\textbf{0.96421}\\
		\cline{2-11}

		&\textit{Baseline+PDBL} &\multirow{3}{*}{LC25000}
&0.93400    &0.96960    &0.97750    &0.98910 &0.98800    &0.99140    &0.99280    &0.99250\\
		&\textit{Baseline*} &
&\textbf{0.95570} &0.97940    &0.98470    &0.99380 &0.99510    &0.99770    &0.99860    &0.99890\\
		&\textit{Baseline*+PDBL} &
&0.95120    &\textbf{0.98080} &\textbf{0.98560} &\textbf{0.99470} &\textbf{0.99550} &\textbf{0.99810} &\textbf{0.99910} &\textbf{0.99950}\\
		\hline
	\end{tabular}\label{tab:quantitative-acc}
\end{table*}

\subsection{Effectiveness of PDBL with Different Proportions of Training Samples}\label{exp:proportions}
In this experiment, we evaluate the effectiveness and efficiency of the proposed PDBL with different proportions of the training set in both datasets. For Kather dataset, Kather Multiclass Internal set~($100k$ patches) and Kather Multiclass External set~($7k$ patches) are the complete training set and test set. For LC25000 dataset, we let 60\% and 40\% samples as the training set and the test set. And then, the training sets of two datasets were randomly split into eight incremental subsets with the proportions of [1\%, 5\%, 10\%, 25\%, 35\%, 50\%, 70\%, 100\%], respectively. We conducted this experiment by comparing three models for every CNN backbone. (1) PDBL directly plugged on the baseline models pre-trained by ImageNet~\cite{deng2009imagenet}, denoted as \textit{Baseline+PDBL}. (2) Baseline models pre-trained by ImageNet fine-tuned for $50$ epochs without PDBL, denoted as \textit{Baseline*}. (3) Baseline models pre-trained by ImageNet fine-tuned for $50$ epochs with PDBL, denoted as \textit{Baseline*+PDBL}. The above notations are used in all the experiments.

Quantitative results for three CNN backbones on two datasets are demonstrated in Table~\ref{tab:quantitative-acc}. First, let us compare the baseline models trained by different training set proportions with and without PDBL (\textit{Baseline*} and \textit{Baseline*+PDBL}). We can observe an overall improvement for nearly all the CNN backbones when equipping with PDBL. With the increasing training samples, the improvements became less significant for three CNN backbones. It can be observed that the degree of the improvements actually depends on three factors, the complexity of the models, the difficulties of the datasets, and the ratio of the training samples. For example, for the deeper backbone with more parameters like ResNet50, the overall improvement on LC25000 is more obvious than Kather. Because LC25000 only has 5 classes but Kather contains 9 classes.
When with enough training samples in LC25000, say 35\%, ResNet50 with PDBL gets only less than 0.001 improvement (0.99510$\rightarrow$0.99550) but it can bring 0.018 improvement (0.97310$\rightarrow$0.99140) for a lightweight backbone ShuffLeNetV2. It is because ResNet50 has a stronger feature representation capability than ShuffLeNetV2, which is a trade-off between computational resources and performance. Despite this, the proposed PDBL is able to steadily improve the classification performance for all the CNN backbones.

Since the proposed PDBL is a model-agnostic plug-and-play module, we are also curious about if semantic features extracted from the backbones trained with ImageNet can be utilized for inference in histopathology images. So for each CNN backbone, we directly applied PDBL on them and updated PDBL for once (\textit{Baseline+PDBL}). In this comparison, different proportions of training samples were only used to update PDBL and did not involve in network training. As can be seen in Table~\ref{tab:quantitative-acc}, baseline models with PDBL can achieve stable classification performances in both datasets comparing with other two competitors (\textit{Baseline*} and \textit{Baseline*+PDBL}). This observation proves that the proposed PDBL can effectively make the inference only relying on the visual semantic features from natural images. For the easier dataset LC25000 with only 5 classes, the performance gap between \textit{Baseline+PDBL} and \textit{Baseline*+PDBL} is relatively small (less than 0.005). For the more difficult dataset Kather with 9 classes, the performance gap between them is around 0.03 accuracy, but the accuracy is still over 0.90. It is interesting that when with smaller training samples (less than 10\%), \textit{Baseline+PDBL} occasionally outperforms \textit{Baseline*} even \textit{Baseline*+PDBL}, especially for the lightweight model ShuffLeNetV2. It is because lack of training samples may lead to overfitting or unstable training problems. For extremely small training data, PDBL provides an alternative solution that uses a well-trained CNN model on ImageNet for feature extraction and to use PDBL for inference. This solution does not need to re-train the network and can greatly save computational resources and annotation efforts.

Based on this experiment, we further discuss the robustness of PDBL with an extremely small training set (only 1\%) in the next experiment.

\begin{table*}[t]
	\centering
	\caption{Robustness of PDBL in an extreme experiment in Kather Dataset~(training set: 1,000 patches; test set: 99,000 patches) and LC25000 Dataset~(training set: 250 patches; test set: 24,750 patches).}
	\begin{tabular}{ll|c|cccccc}
		\hline
		\multicolumn{2}{c|}{\multirow{2}{*}{Models}}& \multirow{2}{*}{Dataset} & \multicolumn{6}{c}{Accuracy}\\
		\cline{4-9}
		&&& Fold1& Fold2& Fold3& Fold4& Fold5&Mean$\pm$SD\\
		\hline
		\multirow{6}{*}{ShuffLeNetV2}&\textit{Baseline+PDBL} &\multirow{3}{*}{Kather}&0.90529 &0.91216 &0.91456 &0.90605 &0.90610&0.90883$\pm$0.004 \\
		&\textit{Baseline*} &
&0.14412 &0.14964 &0.15066 &0.14828 &0.14817&0.14817$\pm$0.002 \\
		&\textit{Baseline*+PDBL} &
&\textbf{0.91164} &\textbf{0.91679} &\textbf{0.91831} &\textbf{0.91517} &\textbf{0.91819}& \textbf{0.91602}$\pm$0.003\\
		\cline{2-9}

		&\textit{Baseline+PDBL} &\multirow{3}{*}{LC25000}
&\textbf{0.95733}    &\textbf{0.95956}    &\textbf{0.95095}    &\textbf{0.95301} &0.95051 &\textbf{0.95427}$\pm$0.004 \\
		&\textit{Baseline*} &
&0.42461    &0.43818    &0.54149    &0.41709 &0.62816 &0.48991$\pm$0.092 \\
		&\textit{Baseline*+PDBL} &
&0.95499 &0.95414 &0.95079 &0.95123 &\textbf{0.95103} &0.95244$\pm$0.002\\
		\hline
		
		\multirow{6}{*}{EfficientNetb0}&\textit{Baseline+PDBL} &\multirow{3}{*}{Kather}
&0.92166 &0.92842 &0.92677 &0.92209 &0.91803 &0.92339$\pm$0.004\\
		&\textit{Baseline*} &
&0.90577 &0.90841 &0.91234 &0.90911 &0.91201 &0.90953$\pm$0.003\\
		&\textit{Baseline*+PDBL} &
&\textbf{0.94894} &\textbf{0.95266} &\textbf{0.95602} &\textbf{0.94843} &\textbf{0.94926} &\textbf{0.95106}$\pm$0.003\\
		\cline{2-9}

		&\textit{Baseline+PDBL} &\multirow{3}{*}{LC25000}
&0.95564    &0.95681    &0.95487    &0.95725 &0.94978 &0.95487$\pm$0.003\\
		&\textit{Baseline*} &
&0.91640    &0.93689    &0.93345    &0.94101 &0.93782 &0.93311$\pm$0.010\\
		&\textit{Baseline*+PDBL} &
&\textbf{0.96230} &\textbf{0.96194} &\textbf{0.95984} &\textbf{0.95960} &\textbf{0.96299} &\textbf{0.96133}$\pm$0.002\\
		\hline

		\multirow{6}{*}{ResNet50}&\textit{Baseline+PDBL} &\multirow{3}{*}{Kather}
&0.92028 &0.92207 &0.92342 &0.91214 &0.91886&0.91935$\pm$0.004\\
		&\textit{Baseline*} &
&0.96063 &0.95318 &0.95742 &0.95080 &0.95435&0.95528$\pm$0.004\\
		&\textit{Baseline*+PDBL} &
&\textbf{0.96078} &\textbf{0.95559} &\textbf{0.95957} &\textbf{0.95207} &\textbf{0.95637}&\textbf{0.95688}$\pm$0.003\\
		\cline{2-9}

		&\textit{Baseline+PDBL} &\multirow{3}{*}{LC25000}
&0.95091    &0.95147    &0.94513    &0.93996 &0.94404 &0.94630$\pm$0.005\\
		&\textit{Baseline*} &
&0.96465    &\textbf{0.96756}    &0.95891    &\textbf{0.96752} &0.96311 &0.96435$\pm$0.004 \\
		&\textit{Baseline*+PDBL} &
&\textbf{0.96537} &0.96642 &\textbf{0.96040} &0.96655 &\textbf{0.96570} &\textbf{0.96489}$\pm$0.003\\
		\hline
	\end{tabular}\label{tab:robustness-acc}
\end{table*}

\subsection{Robustness of PDBL with an extreme proportions of training and test sets.}\label{exp:robustness}
In the previous experiment, we observed a surprising outstanding performance of PDBL with a small training set. In this experiment, we conducted a cross-validation-like experiment, but more difficult than cross-validation, to further discuss the robustness of the proposed PDBL. We randomly sampled 1\% patches (1000 for Kather; 250 for LC25000) from the training data and used the rest 99\% patches (99000 for Kather; 24750 for LC25000) of the training data for testing. To alleviate the sampling bias, we repeated this experiment five times. Each repetition is regarded as a fold in this experiment. All \textit{Baselines*} were pre-trained by ImageNet and fine-tuned for 50 epochs.

According to the accuracy in Table~\ref{tab:robustness-acc}, we notice that the performances of the fine-tuned lightweight models ShuffLeNetV2 and EfficientNetb0 are unstable (\textit{Baseline*}). However, even the quantitative results of ShuffLeNetV2 are nearly random guesses, the proposed PDBL (\textit{Baseline*+PDBL}) can still drastically boost the accuracy from 0.14 to 0.91. Because CNN backbones make an inference by only relying on the high-level semantic features from the last layer. PDBL enriches the features from multiple stages of the CNN backbones, which is beneficial to make an inference when the training samples are very limited.
Furthermore, results of \textit{Baseline*+PDBL} are more stable with smaller standard deviation.

In this experiment, we also directly plugged PDBL on the pre-trained models (\textit{Baseline+PDBL}). Quantitative results prove that the features extracted from the model pre-trained by ImageNet are also stable and effective. For ShuffLeNetV2, it is even higher than
\textit{Baseline*+PDBL}. The reason may be that ShuffLeNetV2 failed to learn a stable model with such limited training samples. Since 1\% training samples of LC25000 is 250, it means that there are only 50 patches for each class. Instead of training an unstable model for histopathology images, PDBL coupling with a more stable model trained from natural images may be an alternative solution for very few training samples.

\subsection{Ablation Study of Pyramidal Design}\label{exp:ablation}
In this experiment, an ablation study was conducted to evaluate the effectiveness of the pyramidal design of the proposed PDBL. We keep the same experimental setting of Section~\ref{exp:robustness} by using 1\% samples for training and 99\% samples for testing five times. Here, we only show the mean and standard deviation of the results. Note that \textit{+DBL} and \textit{+PDBL} represent our method without and with the image pyramid, respectively.

Table~\ref{tab:ablation_acc} and Table~\ref{tab:ablation_f1} demonstrate the accuracy and F1 score on both datasets with only 1\% training samples. We can observe a constantly stable improvement with pyramidal design for all three backbones on both datasets. According to the quantitative results, we found that the improvement of pyramidal design is more significant on the lightweight model. In Kather, PDBL can introduce around 0.09 improvement for ShuffLeNetV2, 0.03 for EfficientNetb0 and only 0.0006 for ResNet50 respectively. In fact, comparing ShuffLeNetV2 with ResNet50, ResNet50 has stronger capacity and feature representation ability, which are great advantages when very few training samples are involved.

In addition, the same finding is also verified in this experiment that pyramidal design brings more improvement for more difficult datasets. For Kather with 9 classes, pyramidal design can introduce multi-level contextual features, which can support the feature representation for more classes.

\begin{table}[t]
	\centering
	\caption{Ablation of Pyramidal Design~(Accuracy)}
    \begin{tabular}{ll|c|c}
		\hline
		\multicolumn{2}{c|}{\multirow{2}{*}{Models}} &\multicolumn{2}{c}{Accuracy~(Mean$\pm$SD)}\\ \cline{3-4}
        &&Kather&LC25000\\ \hline
        \multirow{2}{*}{ShuffLeNetV2}
        &\textit{+DBL}&0.81173$\pm$0.013&0.94772$\pm$0.005\\
        &\textit{+PDBL}&\textbf{0.90883}$\pm$0.004& \textbf{0.95427}$\pm$0.004 \\ \hline
        \multirow{2}{*}{EfficientNetb0}
        &\textit{+DBL}&0.89725$\pm$0.004&0.95175$\pm$0.004\\
        &\textit{+PDBL}&\textbf{0.92339}$\pm$0.004& \textbf{0.95487}$\pm$0.003\\ \hline
        \multirow{2}{*}{ResNet50}
        &\textit{+DBL}&0.91912$\pm$0.004&0.94462$\pm$0.003\\
        &\textit{+PDBL}&\textbf{0.91935}$\pm$0.004& \textbf{0.94630}$\pm$0.005\\ \hline
	\end{tabular}\label{tab:ablation_acc}
\end{table}
\begin{table}[t]
	\centering
	\caption{Ablation of Pyramidal Design~(F1 score)}
    \begin{tabular}{ll|c|c}
		\hline
		\multicolumn{2}{c|}{\multirow{2}{*}{Models}} &\multicolumn{2}{c}{F1~(Mean$\pm$SD)}\\ \cline{3-4}
        &&Kather&LC25000\\ \hline
        \multirow{2}{*}{ShuffLeNetV2}
        &\textit{+DBL}&0.81464$\pm$0.013&0.94793$\pm$0.005\\
        &\textit{+PDBL}&\textbf{0.90947}$\pm$0.004& \textbf{0.95440}$\pm$0.004\\ \hline
        \multirow{2}{*}{EfficientNetb0}
        &\textit{+DBL}&0.89798$\pm$0.004&0.95191$\pm$0.004\\
        &\textit{+PDBL}&\textbf{0.92342}$\pm$0.005& \textbf{0.95496}$\pm$0.003\\ \hline
        \multirow{2}{*}{ResNet50}
        &\textit{+DBL}&0.91927$\pm$0.004&0.94476$\pm$0.003\\
        &\textit{+PDBL}&\textbf{0.91984}$\pm$0.004& \textbf{0.94638}$\pm$0.005\\ \hline
	\end{tabular}\label{tab:ablation_f1}
\end{table}

\begin{table*}[t]
	\centering
	\caption{Domain adaptation from Kather Multiclass Dataset to Zhao~\etal~\cite{ZHAO2020103054}. * means fine-tuning CNN backbones for one epoch. $\dagger$ means updating PDBL for once.}
	\begin{tabular}{c|l|cc|cc|cc}
		\hline
		&\multirow{2}{*}{Models} &\multicolumn{2}{c|}{ShuffLeNetV2} &\multicolumn{2}{c|}{EfficientNetb0}& \multicolumn{2}{c}{ResNet50}\\ \cline{3-8}
		&&Acc &F1 &Acc & F1 &Acc &F1 \\ \hline
		\multirow{2}{*}{No re-training}&\textit{Baseline}
&0.77931 &0.69545  &0.79387 &0.74014  &0.83666 &0.76890 \\
		&\textit{Baseline+PDBL}
&0.74612 &0.67446  &0.79874 &0.75906  &0.85536 &0.80056 \\ \hline
		%%%%%%%%%%%%%%%%%%%%%%%%%%%%%%%%%%%%%%%%%%%%%%%%%%%%%%%%%%%%%%%
		\multirow{3}{*}{1\%}&\textit{Baseline+PDBL$\dagger$}
&\textbf{0.96336} &\textbf{0.95980} &0.97064 & 0.96790 &0.96848 &0.96524 \\
		&\textit{Baseline*}
&0.94626  &0.94032 &0.93037 &0.92082 &0.95875 &0.95416 \\
		&\textit{Baseline*+PDBL$\dagger$}
&0.96275 &0.95910 &\textbf{0.97122} &\textbf{0.96845} &\textbf{0.97183} &\textbf{0.96906} \\ \hline
		%%%%%%%%%%%%%%%%%%%%%%%%%%%%%%%%%%%%%%%%%%%%%%%%%%%%%%%%%%%%%%%
		\multirow{3}{*}{10\%}&\textit{Baseline+PDBL$\dagger$}
&\textbf{0.97952} &\textbf{0.97775} & \textbf{0.98349} & \textbf{0.98203} &0.98190 &0.98031 \\
		&\textit{Baseline*}
&0.96624  &0.96295 &0.96451 &0.96071 &0.97678 &0.97448 \\
		&\textit{Baseline*+PDBL$\dagger$}
&0.97930 &0.97767 &0.98325 &0.98175 &\textbf{0.98454} &\textbf{0.98454} \\ \hline
		%%%%%%%%%%%%%%%%%%%%%%%%%%%%%%%%%%%%%%%%%%%%%%%%%%%%%%%%%%%%%%%
		\multirow{3}{*}{100\%}&\textit{Baseline+PDBL$\dagger$}
&0.98482 &0.98347  &0.98587 &0.98463  &0.98444 &0.98302 \\
		&\textit{Baseline*}
&0.98133 &0.97952 &0.98114 &0.97941 &0.98753 &0.98641 \\
		&\textit{Baseline*+PDBL$\dagger$}
&\textbf{0.98590} &\textbf{0.98463} &\textbf{0.98658} &\textbf{0.98496} &\textbf{0.98987} &\textbf{0.98887} \\ \hline
	\end{tabular}\label{tab:domain adaptation}
\end{table*}
\subsection{Domain Adaptation Study}\label{exp:domain}
Domain adaptation is a crucial ability for a neural network model. In this experiment, we tested if our proposed PDBL can be easily adapted from the source domain to the target domain. The same with Kather Multiclass Dataset, Zhao~\etal~\cite{ZHAO2020103054} also released a histopathological tissue classification dataset for colorectal cancer with night tissue types, which were collected from four different centers, including TCGA, Kather, Guangdong Provincial People's Hospital and Yunnan Cancer Center. We use the data from Guangdong Provincial People's Hospital~($105k$ patches) as the target domain, which were divided into a training set~($63k$~patches) and a test set~($42k$ patches) in this experiment. Let CNN backbones with and without PDBL be trained by the source domain (Kather), we first directly applied them to the target domain~\cite{ZHAO2020103054}. Next, we fine-tuned the models with 100\%~($63k$~patches), 10\%~($6.3k$~patches) and 1\%~($636$ patches) of the training set from the target domain, respectively. (1) The models trained on the source domain Kather Dataset~(100\% KMI) are denoted as $\textit{Baseline}$ and $\textit{Baseline+PDBL}$ in this study. (2) \textit{Baseline*} represents fine-tuning CNN backbones for only one epoch and \textit{+PDBL$\dagger$} means updating PDBL by the training set of target domain. 1\%, 10\%, and 100\% indicate the ratio of training samples we used for fine-tuning CNN backbones and updating PDBL.

Table~\ref{tab:domain adaptation} demonstrates the quantitative results of the domain adaptation study. When we directly apply baseline models (with and without PDBL) trained by Kather to Zhao~\etal~\cite{ZHAO2020103054}, the performance drastically decreases. It means that there exists a domain shift between two datasets. Then we fine-tuned the models by 1\%, 10\%, and 100\% training samples respectively, with only one epoch fine-tuning of baselines and weights updating of PDBL. \textit{Baseline*+PDBL$\dagger$} came back to relatively high performance with the complete training set (100\%). It is interesting that even we do not re-train the baseline models, \textit{Baseline+PDBL$\dagger$} can still obtain an outstanding performance by updating PDBL for once with only 1\% training samples (636 patches). This observation also supports our conclusion in Section~\ref{exp:proportions} that using a more stable baseline model for feature extraction and PDBL for inference is a good solution when training samples are limited. It greatly saves computational resources and annotation efforts.

\subsection{Timing Statistics}
Table~\ref{tab:updating_time} demonstrates the timing statistics of updating weights of PDBL versus training CNN backbones on Kather Multiclass Dataset. The time of one epoch training of CNN models is the average of 50 epochs training time. Since our proposed PDBL only needs to calculate the weights once. According to the timing statistics, the CNN backbones can get a performance boost by only spending around one epoch training time for the calculation of PDBL.
\begin{table}[t]
\centering
\caption{Average training time of baseline models and PDBL on Kather Multiclass Dataset (second).}
\begin{tabular}{c|c|ccc}
\hline
                    &\multirow{2}{*}{\makecell*[c]{CNN\\ (1 epoch)}}&\multicolumn{3}{c}{PDBL~(Total)}\\ \cline{3-5}
                    &&\makecell[c]{Feature\\extraction} &\makecell[c]{Subsequent\\calculation} &Total\\ \hline
    ResNet34        &210    &187   &43     &230\\ \hline
    EfficientNetb0  &345    &394   &85     &479\\ \hline
    ShuffLeNetV2    &157    &180   &33     &213\\ \hline
\end{tabular}\label{tab:updating_time}
\end{table}

\subsection{Semantic Segmentation for Whole Slide Images}
The intention of patch-level tissue classification is to achieve semantic segmentation for whole slide images. So in this experiment, we show a colorectal WSI example of semantic segmentation by our proposed method using the WSI from the department of pathology, Guangdong Provincial People's Hospital. The model is \textit{Baseline*+PDBL} of ResNet50 with 100\% training set in this experiment.

\begin{figure*}[t]\centering
    \setlength{\tabcolsep}{1pt}
    \begin{tabular}{ccccc}
        \includegraphics[width=.33\linewidth]{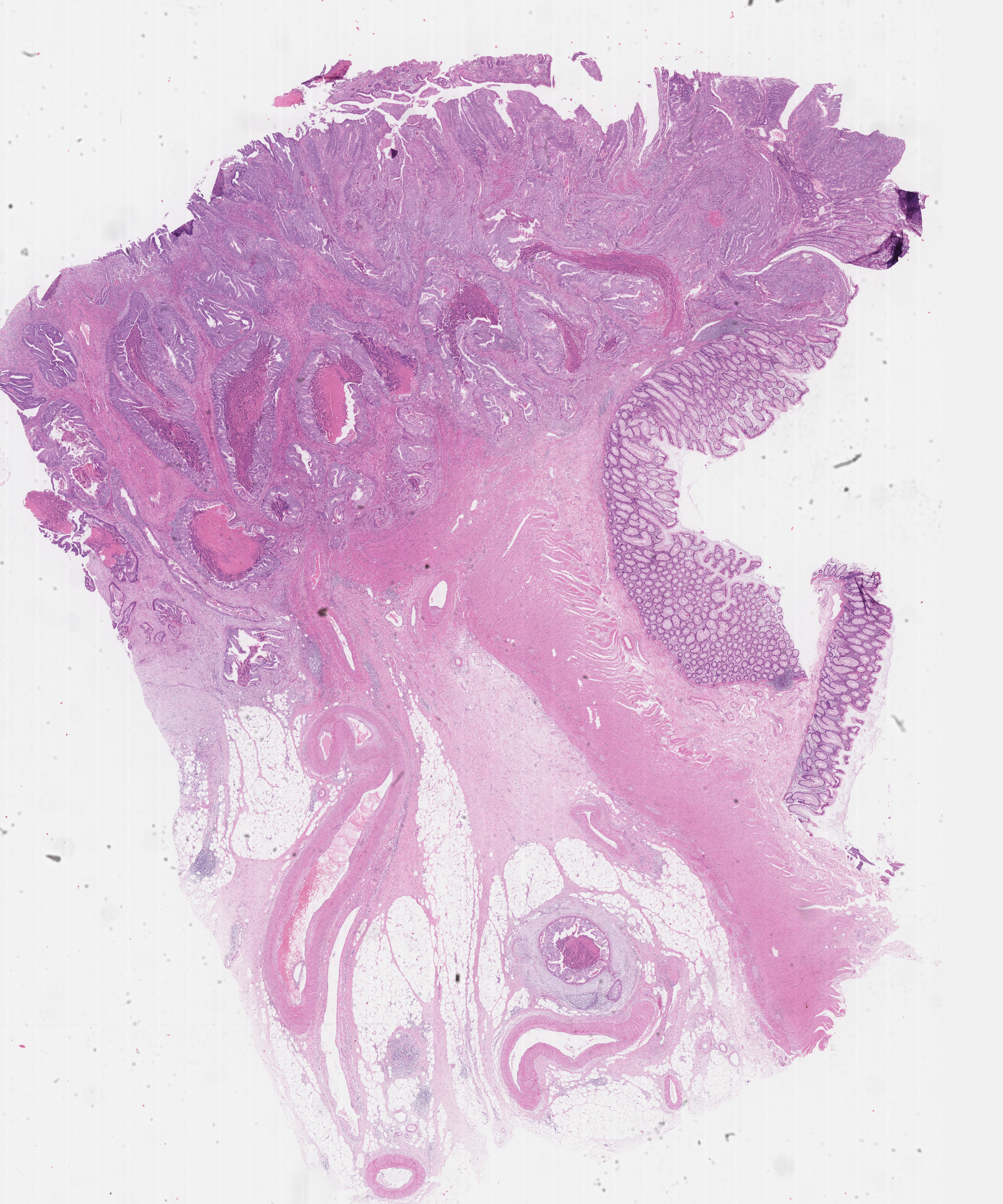}&
        \includegraphics[width=.33\linewidth]{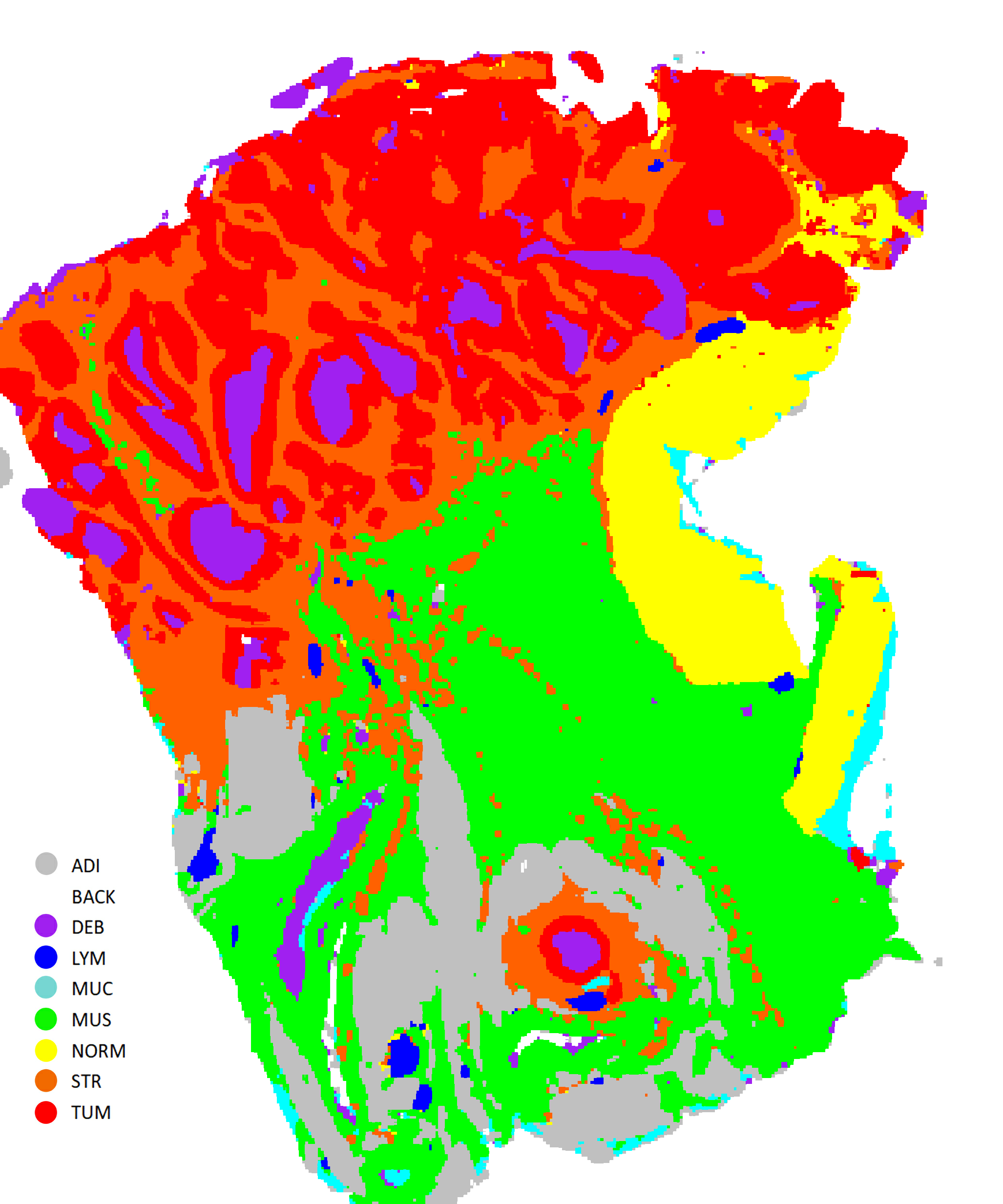}&
        \includegraphics[width=.33\linewidth]{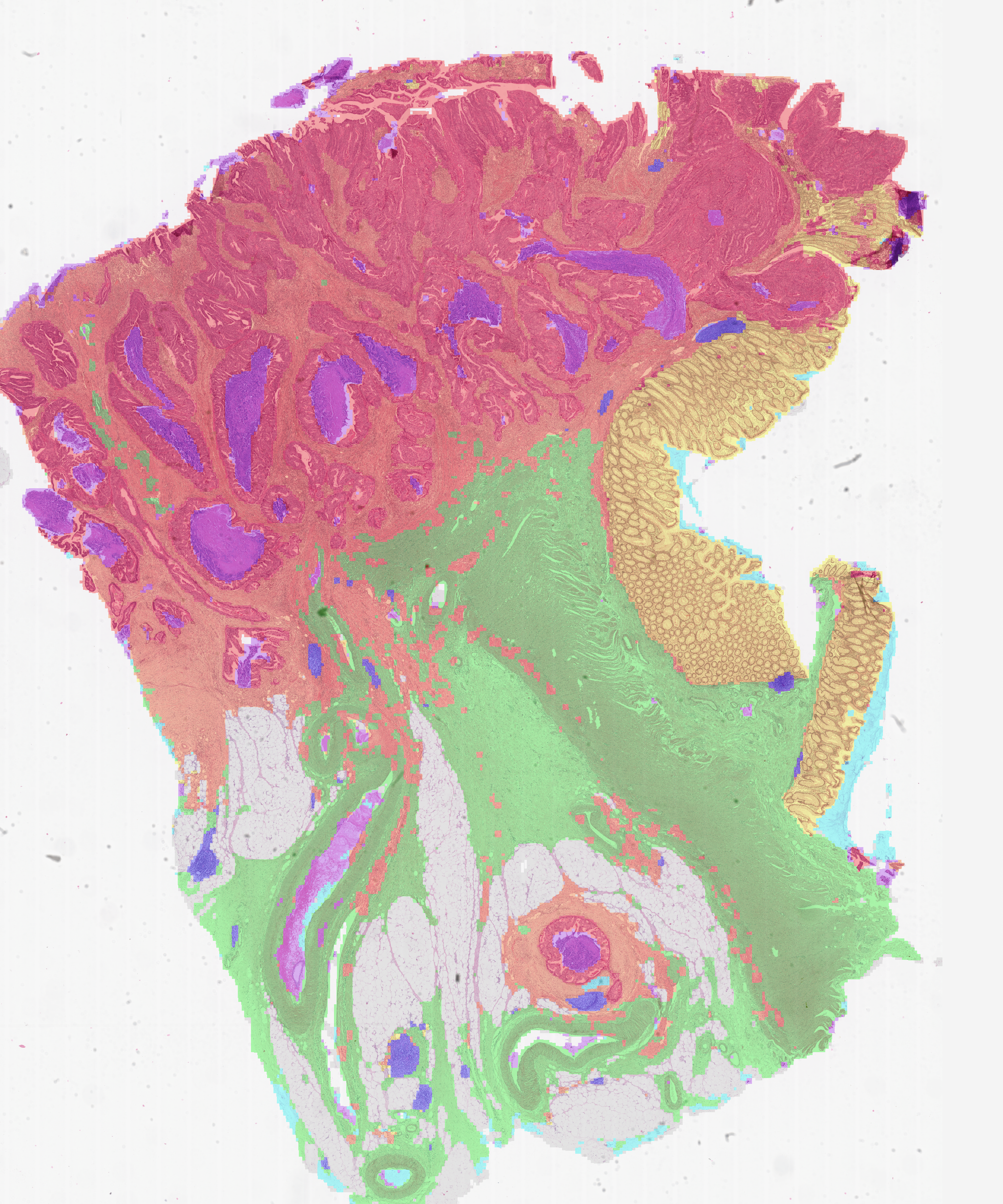}\\
        (a) The original colorectal WSI & (b) Semantic segmentation result & (c) Overlaid result

    \end{tabular}
    \caption{Semantic segmentation results of the colorectal WSI. The original colorectal WSI (a) was cut into 224$\times$224 patches then combined together and formed the semantic segmentation mask (b) and the original image and the mask were overlaid into an overlay (c).}
    \label{fig:seg}
\end{figure*}

Given a WSI in Fig.~\ref{fig:seg}~(a), we first cut it into $224\times 224$ patches under 20$\times$ magnification using sliding windows with the step size of $104$ pixels. For the overlapping region, we decide the tissue class by a voting strategy. So the smaller the step size is, the more the semantic segmentation precision will be, but the more inference time it will spend. Fig.~\ref{fig:seg}~(b) demonstrates the predicted semantic segmentation results. We overlaid the result on the whole slide image for better visualization in Fig.~\ref{fig:seg}~(c).

%!Tex root=./tmi.tex
\section{Conclusion}\label{sec6}
In this paper, we perform histopathological image classification in a new perspective by reconsidering how to make use of the deep features in order to further improve the performance of existing CNN classification backbones. Thus, we proposed a lightweight plug-and-play module called Pyramidal Deep-Broad Learning for any CNN backbone without re-training burden. %We observe that pathologists read pathology slides under different magnifications by keep switching the lens. It inspired us to construct an image pyramid for every input image to obtain pyramidal contextual information. To extract more compact and representative features, we proposed a Deep-Broad block for multi-scale deep-broad feature extraction. PDBL can finally make an inference by fast calculating the weights of the broad learning system. 

We have equipped this plug-and-play module on three representative CNN backbones and achieved a steady improvement of the performance using different proportions of training samples. Specifically, the proposed PDBL demonstrated good feature representation capability and inference ability when very few training samples were involved (less than 10\%), especially for the lightweight models. With PDBL, we even do not have to re-train the baseline models. Such properties can greatly save computational time and annotation efforts. We also look forward to applying this plug-and-play module to more excellent CNN backbones on the other datasets from different tumors in the future. 

%\section*{Acknowledgments}
%
\bibliography{reference}
\bibliographystyle{IEEEtran}

\end{document}